\documentclass[preprint,prd,noshowpacs]{revtex4}
\usepackage{amssymb}
\usepackage{amsmath}

\begin{document}

\title{Hawking-like radiation as tunneling from the apparent horizon in a FRW Universe}

\author{Tao Zhu}
\email{zhut05@lzu.cn}
\author{Ji-Rong Ren}
\email{renjr@lzu.edu.cn}
\affiliation{Institute of Theoretical Physics, Lanzhou University,
Lanzhou 730000, P. R. China}
\author{Douglas Singleton}
\email{dougs@csufresno.edu}
\affiliation{Physics Department, CSU Fresno, Fresno, CA 93740, USA \\
Institute of Gravitation and Cosmology, Peoples' Friendship
University of Russia, Moscow 117198, Russia}

\begin{abstract}
We study Hawking-like radiation in a
Friedmann-Robertson-Walker (FRW) universe using the quasi-classical
WKB/tunneling method which pictures this process as a ``tunneling"
of particles from behind the apparent horizon. The correct
temperature of the Hawking-like radiation from the FRW spacetime is
obtained using a canonical invariant tunneling amplitude. In
contrast to the usual quantum mechanical WKB/tunneling problem where
the tunneling amplitude has only a spatial contribution, we find
that the tunneling amplitude for FRW spacetime (i.e. the imaginary
part of the action) has both spatial and temporal contributions. In
addition we study back reaction and energy conservation of the
radiated particles and find that the tunneling probability and
change in entropy, ${\cal S}$ are related by the relationship:
$\Gamma\propto\exp[-\Delta {\cal S}]$ which differs from the
standard result $\Gamma\propto\exp[\Delta {\cal S}]$. By regarding
the whole FRW universe as an isolated adiabatic system the change in
the total entropy is zero. Then splitting the entropy between
interior and exterior parts of the horizon ($\Delta {\cal S}_{total}
=\Delta {\cal S}_{int} + \Delta {\cal S}_{ext}=0$), we can explain
the origin of the minus sign difference with the usual result: our
$\Delta {\cal S}$ is for the interior region while the standard
result from black hole physics is for the exterior region.
\end{abstract}


\maketitle


\section{Introduction}
\label{secIntroduction}

There has been recent work aimed at establishing the relationship
between the Friedmann equations and the first law of thermodynamics
in the framework of a Friedmann-Robertson-Walker (FRW) universe
\cite{CaiCaoPRD2007,CaiJHEP2005,cai,FRW2007}. If one assumes that
the apparent horizon of a FRW universe has an entropy ${\cal S}=A/4$
and a temperature $T=1/2\pi\tilde{r}_A$, where $A$ and $\tilde{r}_A$
are the area and radius of the apparent horizon, respectively, one
is able to derive the Friedmann equations of the FRW universe with
any spatial curvature by applying the first law of thermodynamics,
$dE=T ~ d {\cal S}$, to the apparent horizon \cite{CaiJHEP2005}.
Other recent work in this direction can be found in
\cite{Zhu2008,Wu,cft/frw,cao2}.

One can derive the Hawking-like temperature, $T=1/2\pi \tilde{r}_A$,
using the first law of thermodynamics and the relationship between
entropy and area of the apparent horizon. In this paper we take a
different starting point for the problem: we first obtain the
temperature using the Hamilton-Jacobi WKB/tunneling method and then
we study the entropy change and back reaction using the first law of
thermodynamics and energy conservation. The WKB method was already
applied to FRW spacetime in a recent paper \cite{CaiHawk08091554},
where the tunneling amplitude for scalar particles across the
apparent horizon of the FRW spacetime was investigated. The same
approach was applied to fermions tunneling from the apparent
horizon\cite{li} and for calculating higher order quantum
corrections \cite{zhu2008Hawking} to the temperature and entropy of
the FRW spacetime. A relationship between the tunneling rate through
the apparent horizon of the FRW spacetime and the first law of
thermodynamics was obtained in \cite{h} using energy conservation.

The tunneling method was initially proposed in \cite{pada5},
\cite{Wilczek}. Due to its simplicity this method has attracted a
lot of attention and subsequent work \cite{other1}. In the tunneling
method of \cite{Wilczek} one calculates the imaginary part of the
action using the null geodesic equation, while \cite{pada5} employs
the Hamilton-Jacobi equations to obtain the particles' classical
action along with detailed balance of the ingoing and outgoing
probability amplitudes. The method of \cite{pada5} has been applied
to more general and complicated spacetimes \cite{other8} and to
dynamical black holes \cite{dynamics}. However, in \cite{canonical},
it was shown that the WKB/tunneling method appeared to give a
temperature twice as large as the correct Hawking temperatures.
Recently, this problem has been solved \cite{time} when it was shown
that in contrast to normal quantum mechanical WKB/tunneling problems
that the tunneling probability received a contribution from the time
coordinate upon crossing the horizon. By requiring canonical
invariance of the tunneling amplitude and taking into account the
temporal contribution one obtains the correct Hawking temperature.
Other recent work in this direction can be found in \cite{2}. In
this paper, we will revisit the Hawking-like radiation from the
apparent horizon of FRW spacetime taking into account these subtle
points. Using these results for the radiation we examine the
relationship between the temperature and the change in entropy of
the interior and exterior of the FRW spacetime. In studying this
connection between temperature and entropy we take into account back
reaction effects and energy conservation of the radiated particle.

The organization of this paper is as follows: In section
\eqref{Hawking} we derive the Hawking-like temperature for a FRW
spacetime using the Hamilton-Jacobi method. If one requires
canonical invariance for the tunneling amplitude one apparently
finds a temperature that is twice the correct value. Upon taking
into account the temporal contribution which occurs from the change
in the time coordinate upon crossing the horizon the correct
temperature is recovered. In section \eqref{back}, we consider the
back reaction effects and energy conservation of the radiated
particles. We also make a connection between the tunneling amplitude
and the entropy change. In section \eqref{conclusion} we give our
conclusions.

\section{Hawking-like Temperature with Canonical Invariance}
\label{Hawking}

The standard form of the FRW metric is
\begin{eqnarray}
\label{frw} ds^2=-dt^2+a^2 (t)
\left(\frac{dr^2}{1-kr^2}+r^2d\Omega_{2}^2 \right),
\end{eqnarray}
where $d\Omega_{2}^2=d\theta^2+sin^2\theta d\varphi^2$ denotes the
line element of an unit two-sphere $S^2$, $a(t)$ is the scale factor
of our universe and $k$ is the spatial curvature constant which can
take values $k=+1$ (positive curvature), $k=0$ (flat), $k=-1$
(negative curvature). In a FRW universe, there is a dynamical
apparent horizon, which is the marginally trapped surface with
vanishing expansion and determined by the
relation\cite{CaiCaoPRD2007}
\begin{eqnarray}
h^{ab}\partial_a\tilde{r}\partial_b\tilde{r}=0,
\end{eqnarray}
where $\tilde{r}=a(t)r$, $h^{ab}=\texttt{diag}(-1,
\frac{1-kr^2}{a^2})$, and $\partial_{a}=(\partial_t,\partial_r)$.
After a simple calculation one can obtain the radius of the apparent
horizon
\begin{eqnarray}
\tilde{r}_A=\frac{1}{\sqrt{H^2+k/a^2}},
\end{eqnarray}
where $H$ is the Hubble parameter, $H\equiv \dot{a}/a$ (the dot
represents derivative with respect to the cosmic time $t$).

In the tunneling approach of reference \cite{Wilczek} the
Painlev\'e-Gulstrand coordinates are used for the Schwarzschild
spacetime. Applying the change of radial coordinate, $\tilde{r}=ar$,
along with the above definitions of $H$ and $\tilde{r}_A$ to the
metric in \eqref{frw} one obtains the Painlev\'e-Gulstrand-like
metric for the FRW spacetime
\begin{eqnarray}
ds^2=-\frac{1-\tilde{r}^2/\tilde{r}_A^2}{1-k
\tilde{r}^2/a^2}dt^2-\frac{2 H \tilde{r}}{1-k \tilde{r}^2/a^2}dt
d\tilde{r}+\frac{1}{1-k\tilde{r}^2/a^2}d\tilde{r}^2+\tilde{r}^2d\Omega_2^2.\label{metric}
\end{eqnarray}
These coordinates have been used in both null geodesic method and
Hamilton-Jacobi method \cite{CaiHawk08091554,li,zhu2008Hawking} to
study the Hawking-like radiation from a FRW metric.

In this paper we will use the Hamilton-Jacobi approach. One starts
by considering a massless scalar field $\phi$ in the FRW spacetime.
This scalar field  obeys the Klein-Gordon equation
\begin{eqnarray}
\frac{-\hbar^2}{\sqrt{-g}}\partial_\mu(g^{\mu\nu}\sqrt{-g}\partial_\nu)\phi=0.\label{kg}
\end{eqnarray}
Substituting the standard ansatz for scalar wave function
$\phi(\tilde{r},t)=\exp[-\frac{i}{\hbar}{\text
S}(\tilde{r},t)+\cdots]$ into \eqref{kg} and taking the limit as
$\hbar\rightarrow 0$, gives the Hamilton-Jacobi equation
\begin{eqnarray}
\label{hj} g^{\mu\nu}\partial_\mu {\text S}\partial_\nu {\text S}=0.
\end{eqnarray}
Since the FRW universe is spherical symmetric, we only need to
examine the $(t-\tilde{r})$ sector of the spacetime. Using the
Painlev\'e-Gulstrand-like coordinates \eqref{metric} in \eqref{hj}
and solving for ${\text S}$, one obtains
\begin{eqnarray}
{\text
S}(\tilde{r},t)=-\int\frac{\omega}{\sqrt{1-k\tilde{r}^2/a^2}}dt+\omega\int
\frac{-H\tilde{r}\pm\sqrt{1-k\tilde{r}^2/a^2}}{(1-\tilde{r}^2/\tilde{r}_A^2)
\sqrt{1-k\tilde{r}^2/a^2}}d\tilde{r},\label{solution}
\end{eqnarray}
where the $+/-$ sign corresponds to the outgoing/ingoing solutions,
respectively. We note that in metric \eqref{metric}, one should use
the Kodama vector \cite{kodama} to define the energy of particle as
$\omega=-\sqrt{1-k\tilde{r}^2/a^2}\partial_{t} S$. Here
$\sqrt{1-k\tilde{r}^2/a^2}\partial_t$ is the Kodama vector. It is
obvious that the action $S$ has a pole at the apparent horizon.
Through a contour integral, we obtain the imaginary part of the
action for both outgoing and ingoing solutions
\begin{eqnarray}
\label{actionout} \texttt{Im} {\text
S}_{\texttt{out}}&=&\texttt{Im}\int\frac{-H\tilde{r}+\sqrt{1-k\tilde{r}^2/a^2}}{(1-\tilde{r}^2/\tilde{r}_A^2)
\sqrt{1-k\tilde{r}^2/a^2}}\omega d\tilde{r}=0,\\
\texttt{Im}{\text S}_{\texttt{in}}&=&
\texttt{Im}\int\frac{-H\tilde{r}-\sqrt{1-k\tilde{r}^2/a^2}}{(1-\tilde{r}^2/\tilde{r}_A^2)
\sqrt{1-k\tilde{r}^2/a^2}}\omega d\tilde{r}=\pi \omega\tilde{r}_A.
\label{actionin}
\end{eqnarray}
The above results are similar to what one finds in the Schwarzschild
case \cite{Wilczek} except now it is the imaginary part of the {\it
outgoing} action which has zero imaginary contribution. In the
Schwarzschild case it was the {\it ingoing} action which had zero
imaginary part. This distinction will be important in the next
section when we relate the change in entropy of the FRW spacetime to
the tunneling amplitude. Note also that unlike the usual quantum
mechanical tunneling problem one finds different contributions to
the imaginary part of the action in these coordinates for the
outgoing versus the ingoing case.

To obtain the temperature one equates the Boltzmann weight,
$\exp[-\omega /T]$, with the tunneling probability $\Gamma$ which is
usually written as
\begin{equation}
\label{gamma} \Gamma\propto\exp[-2\texttt{Im}{\text S}].
\end{equation}
In order to obtain the correct temperature (i.e. $T=\frac{1}{2\pi
\tilde{r}_A}$) one must choose $\texttt{Im} {\text S} =
\texttt{Im}{\text S} _{\texttt{in}}$ rather than $\texttt{Im}{\text
S} = \texttt{Im}{\text S} _{\texttt{out}}$. Aside from having to
make this arbitrary choice in order to obtain the correct result it
has been noted \cite{canonical} that \eqref{gamma} is not invariant
under canonical transformations. One should instead use the
canonically invariant expression
\begin{equation}
\label{gamma1} \Gamma\propto\exp[-\texttt{Im}\oint pdr]
\end{equation}
where $\oint pdr={\text S}_{in}-{\text S}_{out}$. Note that
\eqref{gamma} and \eqref{gamma1} are numerically the same in the
usual case when the action ${\text S}$ has the same magnitude for
the ingoing and outgoing directions crossing the barrier. This is
not the case for the Painlev\'e-Gulstrand-like coordinates of
\eqref{metric}. If one uses \eqref{gamma1} and \eqref{actionout},
\eqref{actionin} to obtain the temperature one finds twice the
temperature -- $T=\frac{1}{\pi \tilde{r}_A}$.

To see the resolution of this problem with the temperature we
transform the Painlev\'e-Gulstrand-like coordinates of
\eqref{metric} via the following change of the time coordinate
\begin{eqnarray}
dt\rightarrow \label{transform}
dt_{*}=dt+\frac{H\tilde{r}}{1-\tilde{r}^2/\tilde{r}_A^2}d\tilde{r}.
\end{eqnarray}
In this way \eqref{metric} takes a Schwarzschild-like form
\begin{eqnarray}
ds^2=-\frac{1-\tilde{r}^2/\tilde{r}_A^2}{1-k\tilde{r}^2/a^2}dt_{*}^2+
\frac{1}{1-\tilde{r}^2/\tilde{r}_A^2}d\tilde{r}^2+\tilde{r}^2d\Omega_2^2.\label{shme}
\end{eqnarray}

For coordinates \eqref{shme}, the Kodama vector is
$K=\sqrt{1-k\tilde{r}^2/a^2}(\partial/\partial_{t_*})$. Thus the
energy of the particle is defined as
\begin{eqnarray}
\omega=-\sqrt{1-k\tilde{r}^2/a^2}\partial_{t_*}S.
\end{eqnarray}
Thus one obtains
\begin{eqnarray}
{\text
S}(\tilde{r},t_*)=-\int\frac{\omega}{\sqrt{1-k\tilde{r}^2/a^2}}dt_*\pm
\omega\int
\frac{1}{1-\tilde{r}^2/\tilde{r}_A^2}d\tilde{r}.\label{solution2}
\end{eqnarray}
The imaginary parts of the action for outgoing and ingoing particles
are
\begin{eqnarray}
\label{action1out} \texttt{Im}{\text
S}_{\texttt{out}}&=&\texttt{Im}\int\frac{1}{1-\tilde{r}^2/\tilde{r}_A^2}\omega
d\tilde{r}=-\frac{\pi
\omega\tilde{r}_A}{2},\\
\texttt{Im}{\text S}_{\texttt{in}}&=&-
\texttt{Im}\int\frac{1}{1-\tilde{r}^2/\tilde{r}_A^2}\omega
d\tilde{r}=\frac{\pi \omega\tilde{r}_A}{2}. \label{action1in}
\end{eqnarray}
In this case regardless of if one takes the tunneling probability as
$\Gamma\propto\texttt{exp}[\mp 2\texttt{Im} {\text S}_{\texttt{in},
\texttt{out}}]$ or $\Gamma\propto\exp[-\texttt{Im}\oint pdr]$ and
$\oint pdr={\text S}_{in}-{\text S}_{out}$ one obtains
$T=1/\pi\tilde{r}_A$, which is the twice of the correct value as was
the case with the Painlev\'e-Gulstrand-like coordinates.

The resolution of this factor of two in the temperature lies in an
overlooked temporal contribution the tunneling amplitude. For a
Schwarzschild black hole \cite{time}, the temporal contribution to
the action is found by changing Schwarzschild coordinates into
Kruskal-Szekeres coordinates and then matching different
Schwarzschild time coordinates across the horizon. Things are a bit
different in the case of the FRW universe. Since for the FRW metric
the metric coefficients depend on both radius and time, there is no
time translation Killing vector field as for the static
Schwarzschild spacetime. For the FRW metric one should use the
Kodama vector in place of the Killing vector. For the FRW spacetime,
the Kodama vector is timelike, null and spacelike for the regions
outside, on, and inside the apparent horizon, respectively. Because
the energy of the particle is defined by the Kodama vector, the
discrepancy of Kodama vector inside and outside the horizon will
effect the temporal part of the action. In terms of the Kodama
vector, the time transformation should be defined by
$\partial_\tau=\sqrt{1-k\tilde{r}^2/a^2}(\partial/\partial_{t_*})$
and $d\tau=dt_*/\sqrt{1-k\tilde{r}^2/a^2}$. Then the metric
\eqref{shme} becomes
\begin{eqnarray}
ds^2=-(1-\tilde{r}^2/\tilde{r}_A^2)d\tau^2+
\frac{1}{1-\tilde{r}^2/\tilde{r}_A^2}d\tilde{r}^2+\tilde{r}^2d\Omega_2^2.
\end{eqnarray}
In order to get the temporal contribution to the imaginary part of
action, one must transform above metric into the
Kruskal-Szekeres-like coordinates. The transformation can be written
in the following form
\begin{eqnarray}
\label{T} T\sim \left\{
        \begin{array}{ll}
          (\tilde{r}_A^2-\tilde{r}^2)^{1/2}\sinh{\frac{\tau}{\tilde{r}_A}}, & \hbox{$\tilde{r}<\tilde{r}_A$;} \\
          (\tilde{r}^2-\tilde{r}_A^2)^{1/2}\cosh{\frac{\tau}{\tilde{r}_A}}, & \hbox{$\tilde{r}>\tilde{r}_A$.}
        \end{array}
      \right.,
\end{eqnarray}
$T$ is the Kruskal-Szekeres-like time coordinate. To match different
patches across the horizon, one needs to rotate the time coordinate
$\tau$ as $\tau\rightarrow\tau-i\pi\tilde{r}_A/2$. Therefore, there
is a temporal contribution to the imaginary part of the total
action, given by $\texttt{Im}(\omega\Delta
\tau^{\texttt{in},\texttt{out}})= - \pi \omega\tilde{r}_A/2$. Note that
unlike the spatial contribution the temporal contribution has the
same sign independent of which direction the horizon is crossed.
Thus, using \eqref{gamma1}, \eqref{action1out}, \eqref{action1in} and
adding the ingoing and outgoing temporal contributions we obtain a
tunneling probability of
\begin{eqnarray}
\Gamma\propto \exp[-2\pi \omega\tilde{r}_A].\label{pure}
\end{eqnarray}
On comparing the tunneling probability with the thermal spectrum
$\Gamma\propto\exp[-\omega/T]$, we recover the correct temperature,
$T=1/2\pi \tilde{r}_A$, associated with the apparent horizon. One
could also use the Painlev\'e-Gulstrand-like coordinates of
\eqref{metric} all the way through the calculation to obtain the
correct temperature, but in this case there will be two temporal
contributions. The first temporal contribution is just that already
given by \eqref{T}. The second comes from the pole in the second
term of the transformation \eqref{transform} between the
Painlev\'e-Gulstrand-like coordinates of \eqref{metric} and the
Schwarzschild-like coordinates of \eqref{shme}. The calculation
works in either coordinate system but is actually a bit more
complicated in the Painlev\'e-Gulstrand-like coordinates.

\section{Back Reaction Effects and Entropy}
\label{back}

In this section we study back reaction effects of the radiated
particle on the FRW spacetime and show the relationship to the
tunneling picture of the previous section.  Many earlier works have
stressed the back reaction effects of the tunneling paradigm in
black hole physics \cite{Back}. In the present section we re-examine
these issues but now taking into account the correct canonically
invariant tunneling probability $\Gamma\propto\exp[-\texttt{Im}\oint
pdr]$ and the temporal contribution to the imaginary part of the
action.

The canonically invariant tunneling probability
$\Gamma\propto\exp[-\texttt{Im}\oint pdr]$, contains two
contributions: from particles crossing the horizon from exterior
region to interior region, and also from interior region to exterior
region. The self-gravitation effects of these two processes should
both be taken into account. In the FRW spacetime, the energy inside
the apparent horizon is defined by a quasi-local mass: the
Misner-Sharp mass, $M=\tilde{r}_A/2$. We note that the tunneling of
a particle from the apparent horizon is an instantaneous process. In
this process one can naturally fix the energy inside the apparent
horizon as the Misner-Sharp mass $M$. Therefore, when the the
radiated particle moves across the apparent horizon, the energy
inside the apparent horizon changes to $M+\omega$ where $\omega$ is
the energy flux which has past through the apparent horizon. Also,
as a result of tunneling, the radius of the apparent horizon,
$\tilde{r}_A$, should change to $\tilde{r}_A+\delta\tilde{r}_A$. In
constructing the first law of thermodynamics on the apparent
horizon, a key point is to calculate the amount of energy crossing
the apparent horizon in an infinitesimal time interval. By using the
Misner-Sharp mass $M$, the energy flux passed through the apparent
horizon is defined as
\begin{eqnarray}
dE=(k^t\partial_tM+k^r\partial_rM)dt=d\tilde{r}_A,
\end{eqnarray}
where $k^{t,r}=(1,-Hr)$ is the (approximate) generator of the
apparent horizon and satisfies
$k^r\partial_r\tilde{r}+k^t\partial_t\tilde{r}=0$. Applying the
above relation to the radiation process and considering that we have
fixed the Misner-Sharp mass $M$, one finds
\begin{eqnarray}
\omega=\int_{M}^{M+\omega}dE=\int_{\tilde{r}_A}^{\tilde{r}_A+\delta
\tilde{r}_A}d\tilde{r}_A=\delta\tilde{r}_A.
\end{eqnarray}
In this case, the imaginary part of the action can be rewritten as
\begin{eqnarray}
\texttt{Im}{\text S}_{\texttt{in}}=\int_0^{\omega}\pi(\tilde{r}_A+\omega')d\omega'=\int_M^{M+\omega}\pi(\tilde{r}_A+\omega')dE,\nonumber\\
\texttt{Im} {\text
S}_{\texttt{out}}=-\int_0^{\omega}\pi(\tilde{r}_A+\omega')d\omega'=-\int_M^{M+\omega}\pi(\tilde{r}_A+\omega')dE.\label{imp}
\end{eqnarray}
(In this section one needs to be careful to distinguish action,
${\text S}$, from entropy ${\cal S}$). In the above expressions, the
temporal contribution to the imaginary part of the action has been
incorporated.

According to the first law of thermodynamics applied to the FRW
universe after the radiation of the particle, we have
\begin{eqnarray}
\frac{1}{2\pi (\tilde{r}_A+\omega)}d {\cal S}^{f}=dE.\label{first}
\end{eqnarray}
Substituting \eqref{first} into \eqref{imp}, one can obtain
\begin{eqnarray}
\texttt{Im}{\text S}_{\texttt{in}}=\int\frac{1}{2}d{\cal S}^f =\frac{1}{2}({\cal S}^f-{\cal S}^i)=\frac{1}{2}\Delta {\cal S}.\nonumber\\
\texttt{Im} {\text S}_{\texttt{out}}=-\int\frac{1}{2}d {\cal
S}^f=-\frac{1}{2}({\cal S}^f-{\cal S}^i)=-\frac{1}{2}\Delta {\cal
S}.\label{imp2}
\end{eqnarray}
Here ${\cal S}^f=\pi(\tilde{r}_A+\omega)^2$ and ${\cal
S}^i=\pi\tilde{r}_A^2$ are the entropy of the apparent horizon after
and before the particle radiation, respectively. Using these
considerations of the self-gravitational effects of the radiated
particle in the FRW universe one can immediately find the tunneling
probability
\begin{eqnarray}
\Gamma\propto\exp[-\Delta {\cal S}]=\exp[-2\pi
\omega(\tilde{r}_A+\frac{\omega}{2})].\label{tun}
\end{eqnarray}
Neglecting the quadratic term in $\omega$, the above result reduces
to \eqref{pure}, the result obtained without considering the
self-gravitation of radiated particle. It should be noted that the
expression \eqref{pure} for the tunneling probability implies a pure
thermal spectrum for the apparent horizon radiation. When back
reaction effects and energy conservation are taken into account,
\eqref{tun} indicates that the radiation spectrum deviates from the
pure thermal spectrum.

Comparing this result to the well known tunneling probability
$\Gamma\propto\exp[\Delta {\cal S}]$, which is obtained in
\cite{Wilczek} for a general, stationary, spherically symmetric
black hole background, we find that there is an additional minus in
\eqref{tun} compared with the result in \cite{Wilczek}. We show here
how these two different results are in fact consistent. In the
Schwarzschild spacetime the observer of the radiation is outside the
horizon. In the FRW universe the observer (i.e. the Kodama observer
who sees the radiation), is inside the apparent horizon. In
Schwarzschild spacetime, the tunneling probability is related to the
change in the entropy of the horizon through particles which tunnel
from {\it inside to outside} the horizon. In contrast for the FRW spacetime
the tunneling probability is related to the
change in the entropy of the apparent horizon via particles which tunnel
from {\it outside to inside} the apparent horizon.

We assume the whole FRW spacetime is an isolated adiabatic system
which can be split into two parts:  the interior and the exterior of
the apparent horizon. The entropy of the interior is just the
entropy of the apparent horizon ${\cal S}_{int}$. We use ${\cal
S}_{ext}$ as the entropy of the exterior of the apparent horizon. As
an adiabatic system, the total entropy of the FRW universe is
conserved, thus
\begin{eqnarray}
\Delta {\cal S}_{int}+\Delta {\cal S}_{ext}=0,
\end{eqnarray}
where $\Delta {\cal S}_{ext}$ is the change of the entropy of the
exterior of the apparent horizon due to the tunneling of a particle
from outside to inside the apparent horizon. Taking this viewpoint,
the tunneling probability can now be written as
$\Gamma\propto\exp[\Delta {\cal S}_{ext}]$, which is then consistent
with the result in \cite{Wilczek}.

\section{Conclusion}
\label{conclusion}

In this paper, we have revisited the derivation of the Hawking-like
radiation from the apparent horizon of a FRW universe. By taking
into account the canonical invariant tunneling probability and the
imaginary parts of both the spatial and temporal contributions of
the tunneling amplitude we recover the correct temperature $T=1/2\pi
\tilde{r}_A$.

We also showed how to obtain the relationship between the tunneling
probability and the change in entropy, $\Gamma\propto\exp[-\Delta
{\cal S}_{int}]$, by taking into account back reaction effects and
energy conservation of the radiated particles. At first glance this
result seems to be wrong by a minus sign as compared to a similar
result for spherically symmetric, static black hole spacetimes (i.e.
$\Gamma\propto\exp[\Delta {\cal S}_{\texttt{BH}}]$ obtained in
\cite{Wilczek}). This apparent inconsistency arises because in the
different spacetimes (FRW versus Schwarzschild) the observer of the
radiation is located on different sides of the (apparent) horizons.
In the FRW spacetime the observer sees radiation which tunnels from
outside to inside the apparent horizon; in the Schwarzschild
spacetime the tunneling is from inside to outside the horizon. If we
regard the whole FRW universe as an isolated adiabatic system and
take into account the total entropy conservation of the FRW
universe, this inconsistency is resolved. In this way, the tunneling
probability for the FRW universe can be written as
$\Gamma\propto\exp[\Delta {\cal S}_{ext}]$ which fits well with the
black hole result. ${\cal S}_{ext}$ is the entropy change of the
exterior of the apparent horizon and satisfies $\Delta {\cal
S}_{int}+\Delta {\cal S}_{ext}=0$.

\section*{Acknowledgements}
Tao Zhu and Ji-Rong Ren are supported by the National Natural
Science Foundation of China and the Cuiying Programme of Lanzhou
University. D.S. is supported by a 2008-2009 Fulbright Scholars
grant. D.S. would like to thank Vitaly Melnikov for the invitation
to research at the Center for Gravitation and Fundamental Metrology
and the Institute of Gravitation and Cosmology at PFUR.

\end{document}